\documentclass[conference]{IEEEtran}
\IEEEoverridecommandlockouts
\usepackage{cite}
\usepackage{amsmath,amssymb,amsfonts}
\usepackage{algorithmic}
\usepackage{caption}
\usepackage{subcaption}
\usepackage{graphicx}
\usepackage{textcomp}
\usepackage{xcolor}
\usepackage{array}
\usepackage{hyperref}
\usepackage{etoolbox}
\usepackage{supertabular}
\usepackage{booktabs}
\usepackage{helvet}
\appto\UrlBreaks{\do\-}
\hypersetup{draft} 

\def\BibTeX{{\rm B\kern-.05em{\sc i\kern-.025em b}\kern-.08em
    T\kern-.1667em\lower.7ex\hbox{E}\kern-.125emX}}
\begin{document}

\title{Lincoln AI Computing Survey (LAICS) and Trends\\
\thanks{Distribution Statement A. Approved for public release. Distribution is un-
limited. This material is based upon work supported by the Under Secretary of
Defense for Research and Engineering under Air Force Contract No. FA8702-
15-D-0001 or FA8702-25-D-B002. Any opinions, findings, conclusions or
recommendations expressed in this material are those of the author(s) and
do not necessarily reflect the views of the Under Secretary of Defense for
Research and Engineering.}
}

\author{\IEEEauthorblockN{Albert Reuther, Peter Michaleas, Michael Jones, \\Vijay Gadepally, and Jeremy Kepner} \\
\IEEEauthorblockA{\textit{MIT Lincoln Laboratory Supercomputing Center} \\
Lexington, MA, USA \\
\{reuther,pmichaleas,michael.jones,vijayg,kepner\}@ll.mit.edu}
}

\maketitle

\begin{abstract}

In the past year, generative AI (GenAI) models have received a tremendous amount of attention, which in turn has increased attention to computing systems for training and inference for GenAI. Hence, an update to this survey is due. This paper is an update of the survey of AI accelerators and processors from past seven years, which is called the Lincoln AI Computing Survey -- LAICS (pronounced ``lace''). This multi-year survey collects and summarizes the current commercial accelerators that have been publicly announced with peak performance and peak power consumption numbers. In the same tradition of past papers of this survey, the performance and power values are plotted on a scatter graph, and a number of dimensions and observations from the trends on this plot are again discussed and analyzed. Market segments are highlighted on the scatter plot, and zoomed plots of each segment are also included. A brief description of each of the new accelerators that have been added in the survey this year is included, and this update features a new categorization of computing architectures that implement each of the accelerators. 

\end{abstract}


\section{Introduction}

In the past 10 years, artificial intelligence and machine learning (AI/ML) has garnered much attention, both in the technical press and in the general media. Starting with deep neural networks (DNNs), then convolutional neural networks (CNNs), and recently generative AI (GenAI), the advances and ensuing exuberance has been energized, in part, by ever-increasing parallel computing capabilities. Much of this capability so far has been delivered to the data center marketplace by Nvidia GPUs, but the sheer size of the AI accelerator market, both in data centers and various embedded ``edge'' applications, has attracted vast amounts of venture funding for startups and internal development funding in established companies. Further, it has brought to light a wave of innovation in computational architectures,  processor-memory interactions, and numerical methods which have been, in many cases, decades in the making. 

Perhaps most notable is the emergence of very large GenAI foundation models, which has driven the recent computational demand for both training and inference. Before these GenAI models, AI accelerators focused on matrix-matrix fused multiply-add operations, while GenAI models have been emphasizing more matrix-vector fused multiply-add operations and high memory bandwidth to compensate for somewhat lower arithmetic intensity~\cite{williams2009roofline}.  For training these very large models many more accelerators are being used simultaneously in a synchronous parallel manner interconnected with very high bandwidth networks: Infiniband, NV-Link, and Converged/Ultra Ethernet. However, this survey continues to focus on the accelerators themselves across a wide range of deployed applications -- from sub-watt to multi-kilowatts -- and on single instances of deployment rather than multiple networked instances. Many new AI accelerators have come to market since the last published iteration of this survey in September 2023, and more insight into the programmability and functionality have been published so another edition of this survey is surely due.  

As in past years, this paper continues to focus on accelerators and processors that are geared toward deep neural networks (DNNs), convolutional neural networks (CNNs), and generative AI as they are quite computationally intense.  
This survey focuses on accelerators and processors for inference for a variety of reasons including that defense and national security AI/ML edge applications rely heavily on inference, though plenty of accelerators in the survey are capable of both inference and training, both computationally and with numerical precisions. 
And we will consider all of the numerical precision types that an accelerator supports, but for most of them, their best inference performance (both computationally and in accuracy) is in int8 and/or fp16/bf16 (IEEE 16-bit floating point or Google's 16-bit brain float) precisions. 

For much of the background of this study, please refer to the previous IEEE-HPEC papers that our team has published~\cite{reuther2023lincoln,reuther2022ai,reuther2021ai,reuther2020survey,reuther2019survey}. The background material in these papers include explanations of the AI ecosystem architecture, the history of the emergence of AI accelerators and accelerators in general, a more detailed explanation of the survey scatter plots, and a discussions of broader observations and trends during each of those years. 

\section{Related Work}

With the amount of research articles and technical press attention that AI accelerators have been given recently, it should not be surprising that quite a number of surveys have been published recently. 
There are many surveys~\cite{lindsey1995survey,liao2001neural,misra2010artificial,sze2017efficient,sze2020efficient,langroudi2019digital,chen2020survey,wang2019deep,khan2020ai,rueckert2020digital,rogers2021academics,sunny2021survey} and other papers that cover various aspects of AI accelerators. For instance, the first paper in this multi-year survey included the peak performance of FPGAs for certain AI models; however, several of the aforementioned surveys cover FPGAs in depth so they are no longer included in this survey. Similarly, early surveys often covered research accelerators that were designed, and sometimes produced and tested, to research various features, design choices, and technologies. However, as more commercial accelerators came to market, these research accelerators became less relevant in discussions about what accelerator to use for a project or to deploy in a data center. Hence, research accelerators are no longer included in this survey.  


This multi-year survey effort and this paper continues to focus on gathering a comprehensive list of AI accelerators with their computational capability, power efficiency, and ultimately the computational effectiveness of utilizing accelerators in embedded and data center applications. Along with this focus, this paper mainly compares neural network accelerators that are useful for government and industrial sensor and data processing applications. A few accelerators and processors that were included in previous years' papers have been left out of this year's survey. They have been dropped because they have been surpassed by new accelerators from the same company, they are no longer offered, or they are no longer relevant to the topic.

\section{Survey of Processors}
\label{sec:survey}

This paper is an update to IEEE-HPEC papers from the past seven years~\cite{reuther2023lincoln,reuther2022ai,reuther2021ai,reuther2020survey,reuther2019survey}. 
This survey continues to cast a wide net to include accelerators and processors for a variety of applications including defense and national security AI/ML edge applications. 
The survey collects information on all of the numerical precision types that an accelerator supports, but for most of them, their best inference performance is in int8 or fp16/bf16, so that is what usually is plotted. This survey gathers peak performance and power information from publicly available materials including research papers, technical trade press, and company benchmarks. We have been gathering peak performance and power because it is the most effective for grouping them into application/deployment categories. 

\begin{figure*}[!bth]
    \centering
    \includegraphics[width=\textwidth]{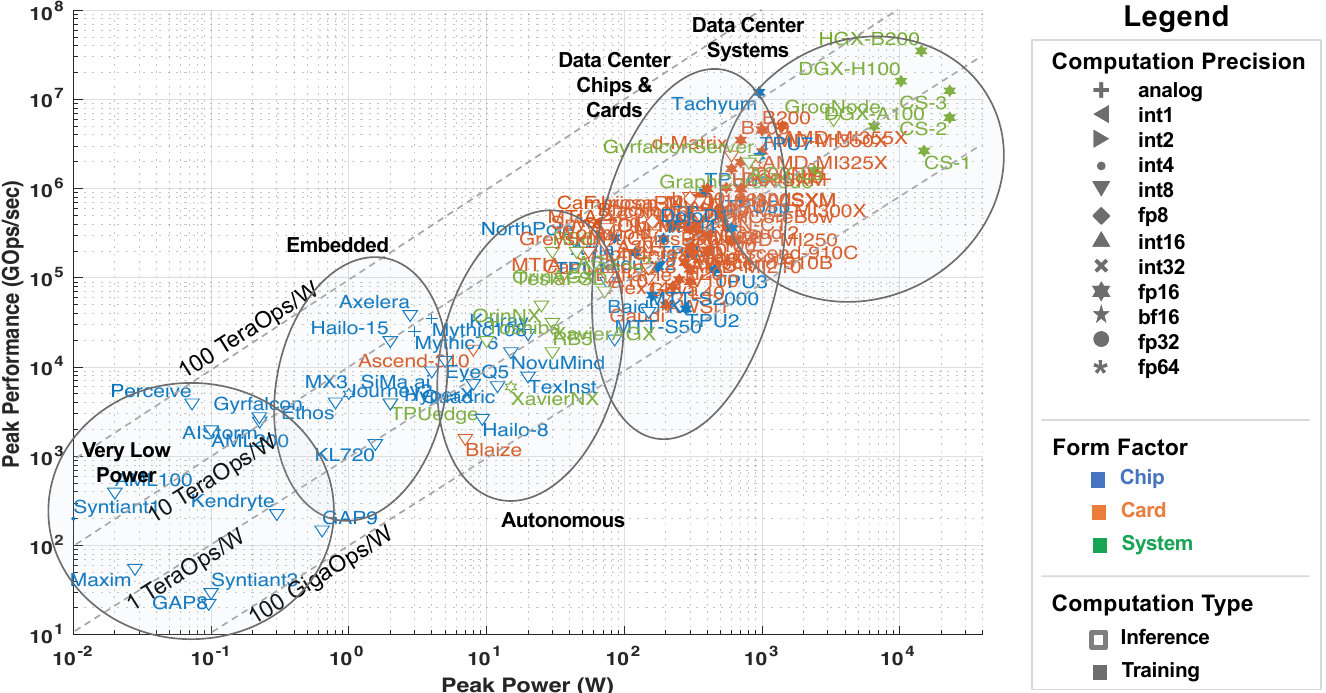}
    \caption{Peak performance vs. power scatter plot of publicly announced AI accelerators and processors.}
    \label{fig:PeakPerformancePower}
  \end{figure*}

\begin{figure*}[!thb]
    \begin{subfigure}{.48\textwidth} %
	\centering
        \includegraphics[width=0.98\textwidth]{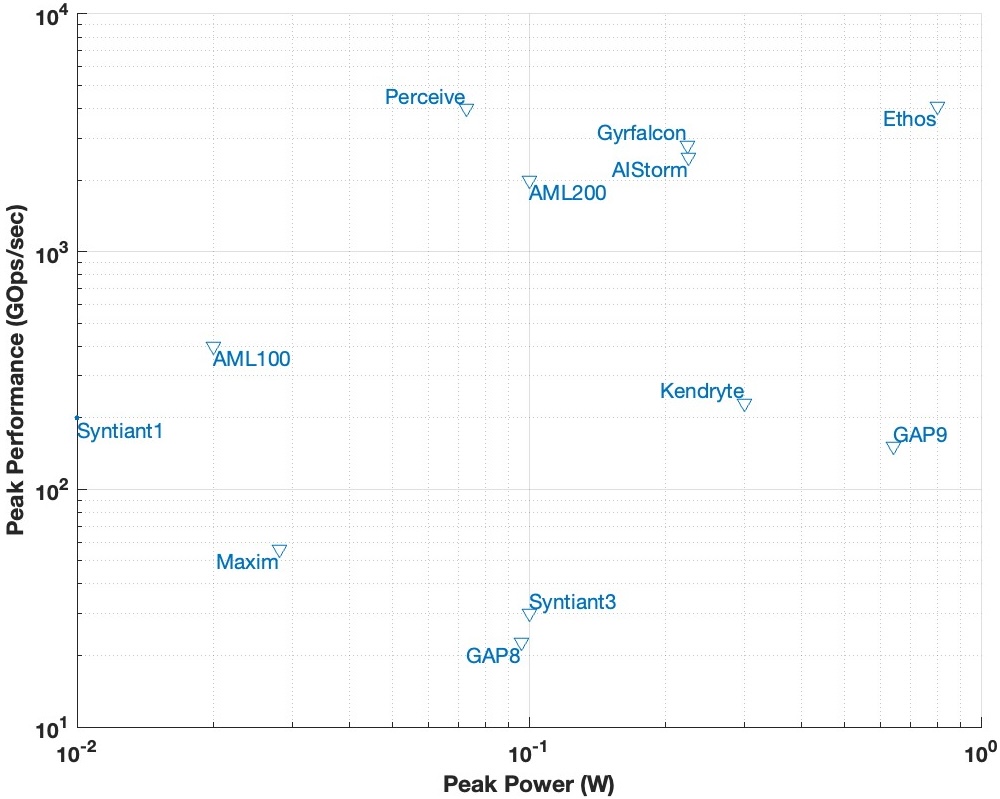}
		\caption{}
    \end{subfigure}
    \begin{subfigure}{.48\textwidth} 
	\centering
        \includegraphics[width=0.98\textwidth]{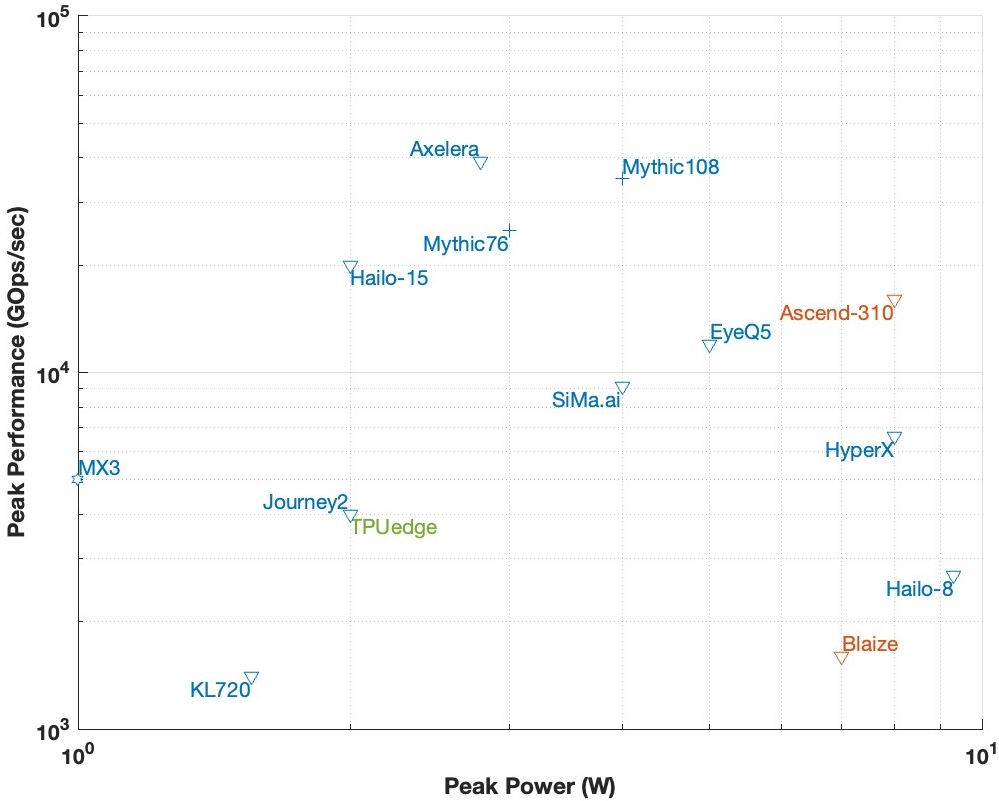}
		\caption{}
    \end{subfigure}
    
    \begin{subfigure}{.48\textwidth} %
	\centering
        \includegraphics[width=0.98\textwidth]{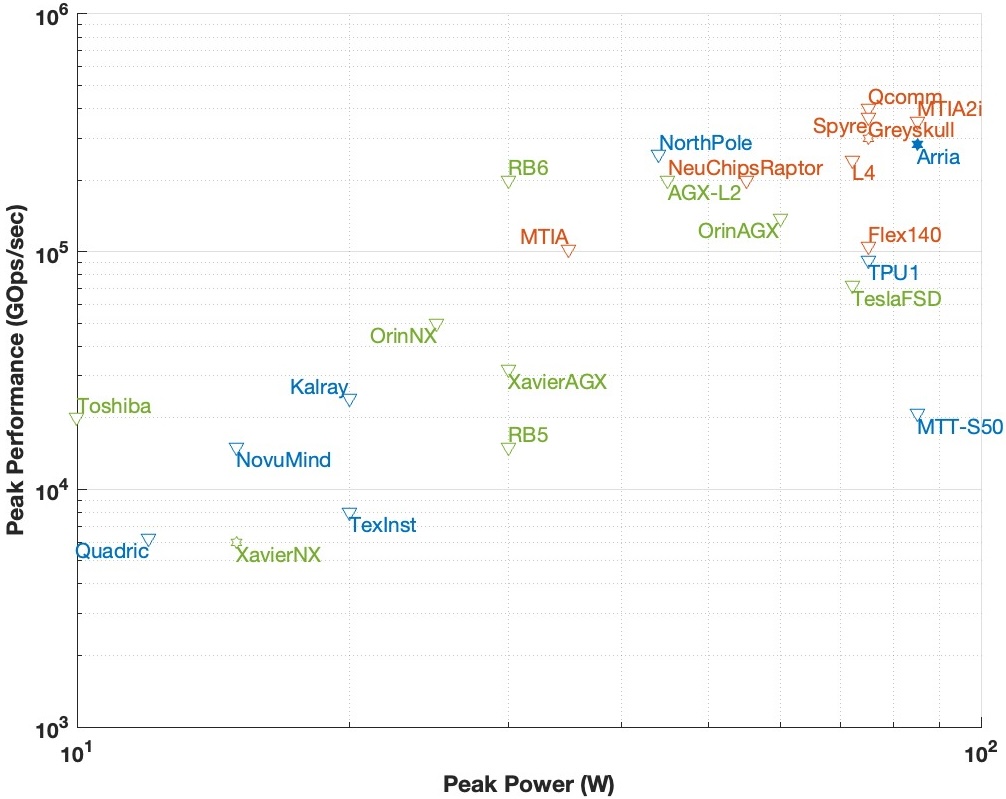}
		\caption{}
    \end{subfigure}
    \begin{subfigure}{.48\textwidth} %
	\centering
        \includegraphics[width=0.98\textwidth]{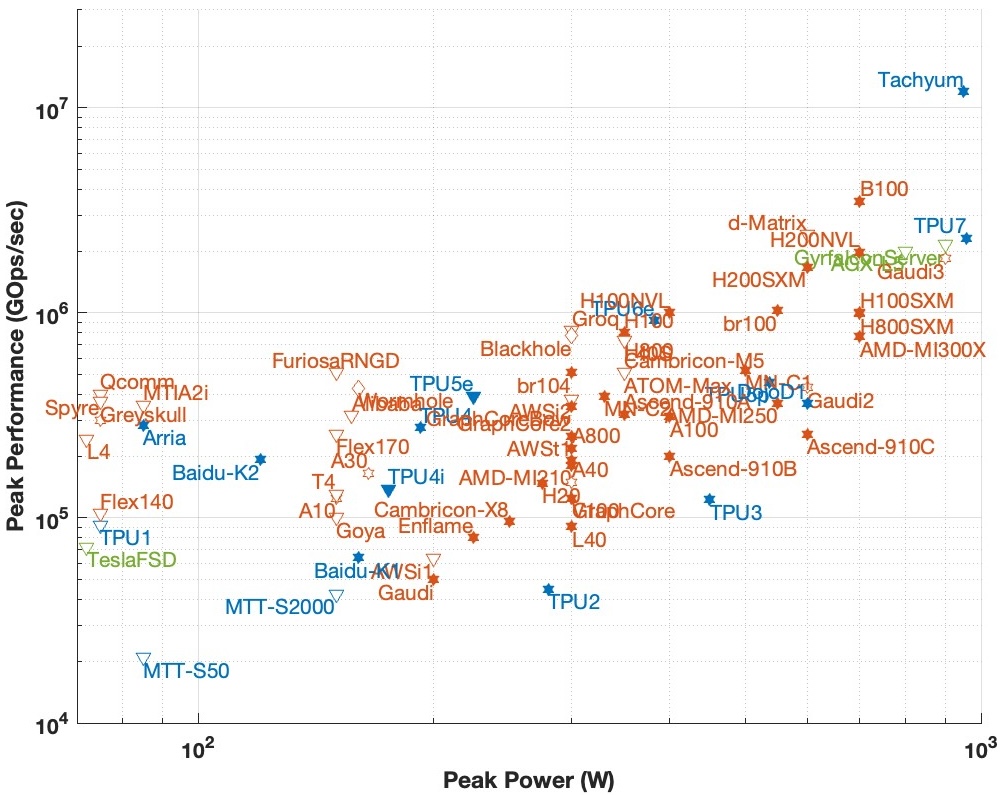}
		\caption{}
    \end{subfigure}

	\centering
    \begin{subfigure}{.48\textwidth} %
        \includegraphics[width=0.98\textwidth]{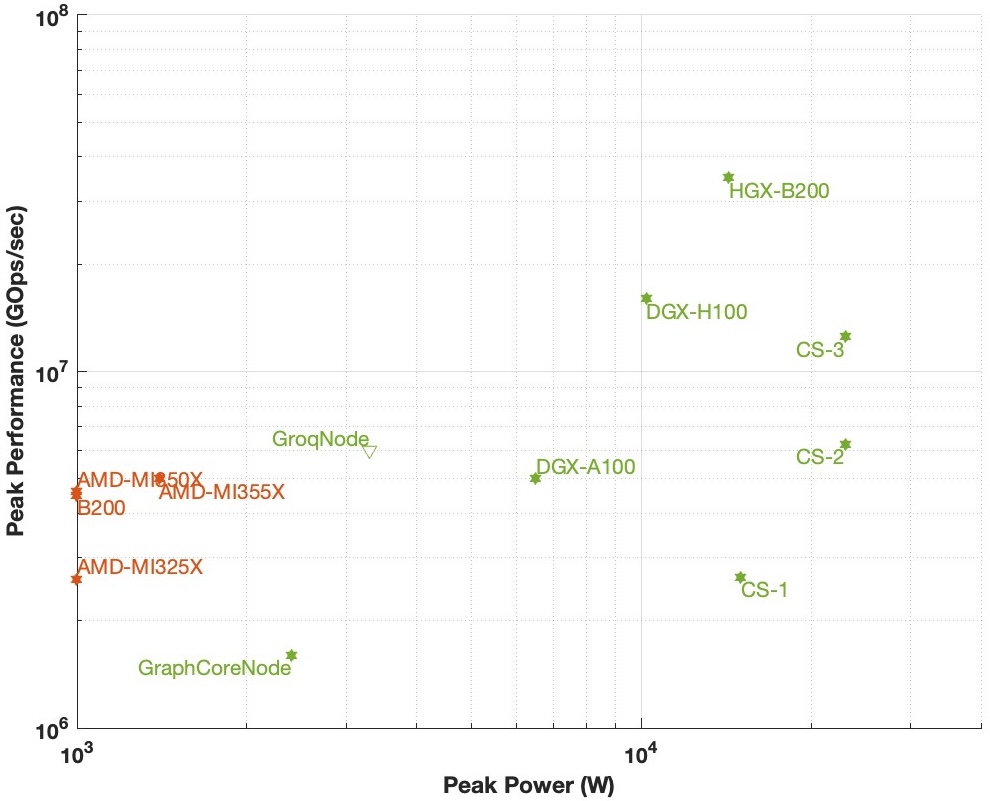}
		\caption{}
    \end{subfigure}

    \caption{Zoomed regions of peak performance vs. peak power scatter plot: \textbf{(a)} very low power, \textbf{(b)} embedded, 
    \textbf{(c)} autonomous, \textbf{(d)} data center chips and cards, \textbf{(e)} data center systems.}
    \label{fig:PeakPerformancePowerZoomed}
\end{figure*}

All of the AI accelerators for which a marker is plotted is listed in Table~\ref{tab:acceleratorlist}, which summarizes some of the important metadata of the accelerators, cards, and systems, including the labels used in Figure~\ref{fig:PeakPerformancePower}. The key metrics of this public data are plotted in Figure~\ref{fig:PeakPerformancePower}, which graphs recent processor capabilities (as of Summer 2025) mapping peak performance vs. peak power consumption.
As in past years, the x-axis indicates peak power, and the y-axis indicate peak giga-operations per second (GOps/s), both on a logarithmic scale. The computational precision of the processing capability is depicted by the marker used. The form factor, for which peak power is reported, is depicted by color: blue corresponds to a single chip; orange corresponds to a card; and green corresponds to entire systems (embedded system, single node desktop and server systems). Finally, the hollow markers are peak performance for inference-only accelerators, while the solid markers are performance for accelerators that are designed to perform both training and inference. The same reasonable categorization of accelerators follows their intended application type. The five categories are shown as ellipses on the graph, which roughly correspond to performance and power consumption: Very Low Power for wake word detection, speech processing, very small sensors, etc.; Embedded for cameras, small UAVs, robots, etc.; Autonomous for driver assist services, autonomous driving, and autonomous robots; Data Center Chips and Cards; and Data Center Systems. A zoomed in scatter plot for each of these categories is shown in the subfigures of Figure~\ref{fig:PeakPerformancePowerZoomed}.

\begingroup
\setlength{\tabcolsep}{2pt} 

  \centering
  \tiny
  
  \tablefirsthead{%
  \textbf{Company} & \textbf{Product} & \textbf{Label} & \textbf{Country} & \textbf{Technology} & \textbf{Form Factor} & \textbf{References} \\ 
     \midrule
}
  \tablehead{%
     \toprule
  \textbf{Company} & \textbf{Product} & \textbf{Label} & \textbf{Country} & \textbf{Technology} & \textbf{Form Factor} & \textbf{References} \\
     \midrule
}
\tabletail{%
 \hline \bottomrule
}
\tablelasttail{%
  \bottomrule 
  }
  \topcaption{List of accelerator metadata, accelerator category, and labels for plots.}
  \label{tab:acceleratorlist}
\begin{supertabular}{@{} l l c c c c c @{}}
     AIStorm & AIStorm & AIStorm & USA & analog & Chip & \cite{merritt2019aistorm}  \\ \hline  
     Alibaba & HanGuang 800 & Alibaba & China & tensor & Card & \cite{peng2019alibaba}  \\ \hline  
     Amazon & Inferentia & AWSi1 & USA & tensor & Card & \cite{morgan2023how}  \\ \hline  
     Amazon & Inferentia2 & AWSi2 & USA & tensor & Card & \cite{morgan2023how}  \\ \hline  
     Amazon & Trainium & AWSt1 & USA & tensor & Card & \cite{morgan2023how}  \\ \hline  
     AMD & MI210 & AMD-MI210 & USA & GPU & Card & \cite{morgan2023third}  \\ \hline  
     AMD & MI250 & AMD-MI250 & USA & GPU & Card & \cite{morgan2023third}  \\ \hline  
     AMD & MI300X & AMD-MI300X & USA & GPU & Card & \cite{morgan2023third}  \\ \hline  
     AMD & MI325X & AMD-MI325X & USA & GPU & Card & \cite{morgan2024amd}  \\ \hline  
     AMD & MI350X & AMD-MI350X & USA & GPU & Card & \cite{alcorn2025amd}  \\ \hline  
     AMD & MI355X & AMD-MI355X & USA & GPU & Card & \cite{alcorn2025amd}  \\ \hline  
     ARM & Ethos N77 & Ethos & UK & tensor & Chip & \cite{schor2020arm}  \\ \hline  
     Aspinity & AML100 & AML100 & USA & analog & Chip & \cite{aspinity2025aml100,hertz2022aspinity}  \\ \hline  
     Aspinity & AML200 & AML200 & USA & analog & Chip & \cite{aspinity2025aml200,hertz2022aspinity}  \\ \hline  
     Axelera & Axelera Test Core & Axelera & Netherlands & manycore & Chip & \cite{ward2022axelera}  \\ \hline  
     Baidu & Baidu Kunlun 200 & Baidu-K1 & China & manycore & Chip & \cite{ouyang2021kunlun,merritt2018baidu,duckett2018baidu}  \\ \hline  
     Baidu & Baidu Kunlun II & Baidu-K2 & China & manycore & Chip & \cite{shilov2021baidu}  \\ \hline  
     Biren Technology & br100 & br100 & China & GPU & Card & \cite{peckham2022chinese,shilov2023chinese,shilov2022chinese}  \\ \hline  
     Biren Technology & br104 & br104 & China & GPU & Card & \cite{peckham2022chinese,shilov2023chinese,shilov2022chinese}  \\ \hline  
     Blaize & El Cano & Blaize & USA & manycore & Card & \cite{demler2020blaize}  \\ \hline  
     Cambricon & MLU290-M5 & Cambricon-M5 & China & GPU & Card & \cite{shilov2025chinas,cambricon2025mlu290}  \\ \hline  
     Cambricon & MLU370-X8 & Cambricon-X8 & China & GPU & Card & \cite{shilov2025chinas,cambricon2025mlu370}  \\ \hline  
     Canaan & Kendrite K210 & Kendryte & Singapore & CPU & Chip & \cite{gwennap2019kendryte}  \\ \hline  
     Cerebras & CS-1 & CS-1 & USA & manycore & System & \cite{hock2019introducing}  \\ \hline  
     Cerebras & CS-2 & CS-2 & USA & manycore & System & \cite{trader2021cerebras}  \\ \hline  
     Cerebras & CS-3 & CS-3 & USA & manycore & System & \cite{morgan2024cerebras}  \\ \hline  
     HyperX Logic & HyperX & HyperX & USA & manycore & Chip & \cite{demler2020coherent}  \\ \hline  
     d-Matrix & Corsair & d-Matrix & USA & manycore & Card & \cite{ward2025d-matrix}  \\ \hline  
     Enflame & Cloudblazer T10 & Enflame & China & CPU & Card & \cite{clarke2019globalfoundries}  \\ \hline  
     FuriosaAI & RNGD & FuriosaRNGD & S. Korea & tensor & Card & \cite{lam2024furiosaai,kim2024tcp}  \\ \hline  
     Google & TPU Edge & TPUedge & USA & tensor & System & \cite{tpu2019edge}  \\ \hline  
     Google & TPU1 & TPU1 & USA & tensor & Chip & \cite{jouppi2020domain,teich2018tearing}  \\ \hline  
     Google & TPU2 & TPU2 & USA & tensor & Chip & \cite{jouppi2020domain,teich2018tearing}  \\ \hline  
     Google & TPU3 & TPU3 & USA & tensor & Chip & \cite{jouppi2021ten,jouppi2020domain,teich2018tearing}  \\ \hline  
     Google & TPU4i & TPU4i & USA & tensor & Chip & \cite{jouppi2021ten}  \\ \hline  
     Google & TPU4 & TPU4 & USA & tensor & Chip & \cite{peckham2022google}  \\ \hline  
     Google & TPU5e & TPU5e & USA & tensor & Chip & \cite{morgan2025with}  \\ \hline  
     Google & TPU5p & TPU5p & USA & tensor & Chip & \cite{morgan2025with}  \\ \hline  
     Google & TPU6e & TPU6e & USA & tensor & Chip & \cite{morgan2025with}  \\ \hline  
     Google & TPU7 & TPU7 & USA & tensor & Chip & \cite{morgan2025with}  \\ \hline  
     GraphCore & C2 & GraphCore & UK & manycore & Card & \cite{gwennap2020groq,lacey2017preliminary}  \\ \hline  
     GraphCore & C2 & GraphCoreNode & UK & manycore & System & \cite{graphcore2020dell}  \\ \hline  
     GraphCore & Colossus Mk2 & GraphCore2 & UK & manycore & Card & \cite{ward2020graphcore}  \\ \hline  
     GraphCore & Bow-2000 & GraphCoreBow & UK & manycore & Card & \cite{tyson2022graphcore}  \\ \hline  
     GreenWaves & GAP8 & GAP8 & France & manycore & Chip & \cite{greenwaves2020gap,turley2020gap9}  \\ \hline  
     GreenWaves & GAP9 & GAP9 & France & manycore & Chip & \cite{greenwaves2020gap,turley2020gap9}  \\ \hline  
     Groq & Groq Node & GroqNode & USA & tensor & System & \cite{hemsoth2020groq}  \\ \hline  
     Groq & Tensor Streaming Proc. & Groq & USA & tensor & Card & \cite{gwennap2020groq,abts2020think}  \\ \hline  
     Gyrfalcon & Gyrfalcon & Gyrfalcon & USA & manycore & Chip & \cite{ward2019gyrfalcon}  \\ \hline  
     Gyrfalcon & Gyrfalcon & GyrfalconServer & USA & manycore & System & \cite{hpcwire2020solidrun}  \\ \hline  
     Hailo & Hailo-8 & Hailo-8 & Israel & manycore & Chip & \cite{ward2019details}  \\ \hline  
     Hailo & Hailo-15H & Hailo-15 & Israel & manycore & Chip & \cite{ward2023hailo}  \\ \hline  
     Horizon Robotics & Journey2 & Journey2 & China & tensor & Chip & \cite{horizon2020journey}  \\ \hline  
     Huawei HiSilicon & Ascend 310 & Ascend-310 & China & manycore & Card & \cite{huawei2020ascend310}  \\ \hline  
     Huawei HiSilicon & Ascend 910A & Ascend-910A & China & manycore & Card & \cite{morgan2024huawei,shilov2025huawei}  \\ \hline  
     Huawei HiSilicon & Ascend 910B & Ascend-910B & China & manycore & Card & \cite{morgan2024huawei,shilov2025huawei}  \\ \hline  
     Huawei HiSilicon & Ascend 910C & Ascend-910C & China & manycore & Card & \cite{morgan2024huawei,shilov2025huawei}  \\ \hline  
     IBM & NorthPole & NorthPole & USA & manycore & Chip & \cite{cassidy2024ibm,akopyan2024breakthrough,appuswamy2024breakthrough}  \\ \hline  
     IBM & Spyre AIU & Spyre & USA & manycore & Card & \cite{morgan2024ibm,berry2024ibm}  \\ \hline  
     Intel & Arria 10 1150 & Arria & USA & FPGA & Chip & \cite{abdelfattah2018dla,hemsoth2018intel}  \\ \hline  
     Intel & Mobileye EyeQ5 & EyeQ5 & Israel & manycore & Chip & \cite{demler2020blaize}  \\ \hline  
     Intel & Flex140 & Flex140 & USA & GPU & Card & \cite{morgan2022different}  \\ \hline  
     Intel & Flex170 & Flex170 & USA & GPU & Card & \cite{morgan2022different}  \\ \hline  
     Intel Habana & Gaudi & Gaudi & Israel & tensor & Card & \cite{gwennap2019habanagaudi,medina2020habana}  \\ \hline  
     Intel Habana & Goya HL-1000 & Goya & Israel & tensor & Card & \cite{gwennap2019habanagoya,medina2020habana,gwennap2019habanagaudi}  \\ \hline  
     Intel Habana & Gaudi2 & Gaudi2 & Israel & tensor & Card & \cite{smith2024intel,morgan2022intel}  \\ \hline  
     Intel Habana & Gaudi3 & Gaudi3 & Israel & tensor & Card & \cite{smith2024intel}  \\ \hline  
     Kalray & Coolidge & Kalray & France & manycore & Chip & \cite{dupont2019kalray, clarke2020nxp}  \\ \hline  
     Kneron & KL720 & KL720 & USA & tensor & Chip & \cite{ward2021kneron}  \\ \hline  
     Maxim & Max 78000 & Maxim & USA & tensor & Chip & \cite{ward2020maxim,jani2021maxim,clay2022benchmarking}  \\ \hline  
     MemryX & MX3 & MX3 & USA & manycore & Chip & \cite{leibson2023adding}  \\ \hline  
     Meta/Facebook & MTIA & MTIA & USA & manycore & Card & \cite{morgan2023meta,firoozshahian2023mtia}  \\ \hline  
     Meta/Facebook & MTIA2i & MTIA2i & USA & manycore & Card & \cite{coburn2025meta,maddury2024next}  \\ \hline  
     Moore Threads & MTT S50 & MTT-S50 & China & GPU & Chip & \cite{klotz2024new}  \\ \hline  
     Moore Threads & MTT S2000 & MTT-S2000 & China & GPU & Chip & \cite{klotz2024new}  \\ \hline  
     Mythic & M1076 & Mythic76 & USA & analog & Chip & \cite{ward2021mythic,hemsoth2018mythic,fick2018mythic}  \\ \hline  
     Mythic & M1108 & Mythic108 & USA & analog & Chip & \cite{ward2021mythic,hemsoth2018mythic,fick2018mythic}  \\ \hline  
     Neuchips & Raptor & NeuChipsRaptor & Taiwan & tensor & Card & \cite{smith2024neuchips}  \\ \hline  
     NovuMind & NovuTensor & NovuMind & USA & tensor & Chip & \cite{freund2019novumind,yoshida2018novumind}  \\ \hline  
     NVIDIA & Ampere A10 & A10 & USA & GPU & Card & \cite{morgan2021nvidia}  \\ \hline  
     NVIDIA & Ampere A100 & A100 & USA & GPU & Card & \cite{krashinsky2020nvidia}  \\ \hline  
     NVIDIA & Ampere A800 & A800 & USA & GPU & Card & \cite{shilov2023nvidias}  \\ \hline  
     NVIDIA & Ampere A30 & A30 & USA & GPU & Card & \cite{morgan2021nvidia}  \\ \hline  
     NVIDIA & Ampere A40 & A40 & USA & GPU & Card & \cite{morgan2021nvidia}  \\ \hline  
     NVIDIA & Broadwell & B100 & USA & GPU & Card & \cite{morgan2024how}  \\ \hline  
     NVIDIA & Broadwell & B200 & USA & GPU & Card & \cite{morgan2024how}  \\ \hline  
     NVIDIA & DGX-A100 & DGX-A100 & USA & GPU & System & \cite{campa2020defining}  \\ \hline  
     NVIDIA & DGX-H100 & DGX-H100 & USA & GPU & System & \cite{mujtaba2022nvidia}  \\ \hline  
     NVIDIA & HGX-B200 & HGX-B200 & USA & GPU & System & \cite{nvidia2025hgx}  \\ \hline  
     NVIDIA & Hopper H100 PCIe & H100 & USA & GPU & Card & \cite{morgan2023nvidias}  \\ \hline  
     NVIDIA & Hopper H100 SXM & H100SXM & USA & GPU & Card & \cite{smith2022nvidia}  \\ \hline  
     NVIDIA & Hopper H100 NVL & H100NVL & USA & GPU & Card & \cite{morgan2023nvidias}  \\ \hline  
     NVIDIA & H800 SXM & H800SXM & USA & GPU & Card & \cite{morgan2025separate}  \\ \hline  
     NVIDIA & H800 SXM PCIe & H800 & USA & GPU & Card & \cite{morgan2025separate}  \\ \hline  
     NVIDIA & Hopper H200 SMX & H200SXM & USA & GPU & Card & \cite{nvidia2025h200}  \\ \hline  
     NVIDIA & Hopper H200 NVL & H200NVL & USA & GPU & Card & \cite{nvidia2025h200}  \\ \hline  
     NVIDIA & H20 & H20 & USA & GPU & Card & \cite{morgan2025separate}  \\ \hline  
     NVIDIA & Jetson AGX Xavier & XavierAGX & USA & GPU & System & \cite{smith2019nvidia}  \\ \hline  
     NVIDIA & Jetson NX Orin & OrinNX & USA & GPU & System & \cite{funk2022nvidia,nvidia2022embedded}  \\ \hline  
     NVIDIA & Jetson AGX Orin & OrinAGX & USA & GPU & System & \cite{funk2022nvidia,nvidia2022embedded}  \\ \hline  
     NVIDIA & Jetson Xavier NX & XavierNX & USA & GPU & System & \cite{smith2019nvidia}  \\ \hline  
     NVIDIA & DRIVE AGX L2 & AGX-L2 & USA & GPU & System & \cite{hill2020nvidia}  \\ \hline  
     NVIDIA & DRIVE AGX L5 & AGX-L5 & USA & GPU & System & \cite{hill2020nvidia}  \\ \hline  
     NVIDIA & L4 & L4 & USA & GPU & Card & \cite{morgan2023nvidias}  \\ \hline  
     NVIDIA & L40 & L40 & USA & GPU & Card & \cite{techpowerup2023nvidia}  \\ \hline  
     NVIDIA & L40S & L40S & USA & GPU & Card & \cite{robinson2023nvidia}  \\ \hline  
     NVIDIA & T4 & T4 & USA & GPU & Card & \cite{kilgariff2018nvidia}  \\ \hline  
     NVIDIA & Volta V100 & V100 & USA & GPU & Card & \cite{volta2019nvidia,smith201816gb}  \\ \hline  
     Perceive & Ergo & Perceive & USA & tensor & Chip & \cite{mcgregor2020perceive}  \\ \hline  
     Preferred Networks & MN-Core1 & MN-C1 & Japan & manycore & Card & \cite{preferred2020mncore,cutress2019preferred,namura2021mn-core}  \\ \hline  
     Preferred Networks & MN-Core2 & MN-C2 & Japan & manycore & Card & \cite{preferred2020mncore,makino2024mn-core2}  \\ \hline  
     Quadric & q1-64 & Quadric & USA & manycore & Chip & \cite{firu2019quadric}  \\ \hline  
     Qualcomm & Cloud AI 100 & Qcomm & USA & GPU & Card & \cite{ward2020qualcomm,mcgrath2019qualcomm}  \\ \hline  
     Qualcomm & QRB5165 & RB5 & USA & GPU & System & \cite{crowe2020qualcomm}  \\ \hline  
     Qualcomm & QRB5165N & RB6 & USA & GPU & System & \cite{qualcomm2023robotics}  \\ \hline  
     Rebellions & ATOM Max & ATOM-Max & S. Korea & tensor & Card & \cite{cozma2024rebellions,ward2024rebellions,hong2025performance}  \\ \hline  
     SiMa.ai & SiMa.ai & SiMa.ai & USA & tensor & Chip & \cite{gwennap2020machine}  \\ \hline  
     Syntiant & NDP101 & Syntiant1 & USA & manycore & Chip & \cite{mcgrath2018tech,merritt2018syntiant}  \\ \hline  
     Syntiant & NDP250 & Syntiant3 & USA & manycore & Chip & \cite{ward2024syntiant}  \\ \hline  
     Tachyum & Prodigy & Tachyum & USA & manycore & Chip & \cite{shilov2022tachyum}  \\ \hline  
     Tenstorrent & Greyskull & Greyskull & USA & manycore & Card & \cite{gwennap2020tenstorrent}  \\ \hline  
     Tenstorrent & Wormhole n300 & Wormhole & USA & manycore & Card & \cite{Ignjatovic2022wormhole,shilov2024tenstorrent}  \\ \hline  
     Tenstorrent & Blackhole & Blackhole & USA & manycore & Card & \cite{mann2024tenstorrent,tenstorrent2025blackhole}  \\ \hline  
     Tesla & Full Self-Driving Comp. & TeslaFSD & USA & tensor & System & \cite{talpes2020compute,wikichip2020fsd}  \\ \hline  
     Tesla & Dojo D1 & DojoD1 & USA & manycore & Chip & \cite{talpes2023microarchitecture,morgan2022inside}  \\ \hline  
     Texas Instruments & TDA4VM & TexInst & USA & manycore & Chip & \cite{ward2020ti,ti2021tda4vm,demler2020ti}  \\ \hline  
     Toshiba & 2015 & Toshiba & Japan & manycore & System & \cite{merritt2019samsung}  \\ \hline  
  
\end{supertabular}
\endgroup

For most of the accelerators, their descriptions and commentaries have not changed since last year so please refer to the previous papers of this survey project for descriptions and commentaries. Several new releases and a few departures are included in this update, and they are chronicled next. 

\begin{itemize} 

\item Amazon AWS has published much more information about their Inferentia and Trainium chips, which have been design by their in-house Annapurna Labs Division. These accelerators are multi-core tensor accelerators~\cite{morgan2023third}, and we can expect newer generations of these chips in coming years. 

\item During the past two years, AMD released the MI300 series of GPUs to compete head-to-head with Nvidia's data center GPUs, namely the MI300X, MI325X, MI350X, and MI355X. Each provided ample competition to the Nvidia counterparts~\cite{alcorn2025amd}, with FP6 and FP4 support and enhanced matrix engines. 

\item Nvidia continues to release a variety of data center GPUs in order to keep their lead in supplying GPUs for both training and inference. In the past two years, Nvidia released the L4 and  L40S for cloud graphics and low-power inference, several variants of the Hopper GPUs (H100 NVL, H200 SMX, and H200 NVL), and two variants of the Broadwell GPU (B100 and B200). The Hopper and Broadwell GPUs are true transformer/generative AI accelerators with larger matrix engines (TensorCores) along with FP6 and FP4 support. To address export restrictions, Nvidia also released detuned versions of their A100, H100, and H200 GPUs with the A800, H800, and H20 GPUs, respectively. 

\item Intel has been consolidating their AI accelerator efforts. Intel's Habana subsidiary released their Gaudi2 and Gaudi3 tensor accelerators, which have gained some good traction. Given that traction, Intel has cancelled their GPU training offerings (Xe-HPC, codename Ponte Vecchio and future Rialto Bridge GPU),  while still offering the Flex line of data center inference GPUs. On the software side, Intel has integrated CPU, GPU, and FPGA programming into their OneAPI software stack. 

\item Alphabet Google has continued to refine and improve the performance of their cloud data center TPUs with the releases of their TPU5e, TPU5p, TPU6e, and TPU7~\cite{morgan2025with}. Each improves performance over the previous generation for their own and their clients' workloads. 

\item Cerebras has released their third generation Wafer Scale Engine (WSE) accelerator, CS-3, with impressive performance both as a single node and networked multi-node training~\cite{morgan2024cerebras}. 

\item Tesla Motors released details about their Dojo1 data center training chip and system,  disclosing an impressive design for very large scale training~\cite{talpes2023microarchitecture,morgan2022inside}. 

\item Meta/Facebook has also released many more details about the first and second generation Meta Training and Inference Accelerator (MTIA and MTIA2i), which are both focused on inference. They are both comprised of an 8-by-8 network meshed set of processing elements, and each processing element includes two RISC-V cores for computation. One of the two RISC-V cores includes a 64-element RISC-V vector extension~\cite{firoozshahian2023mtia,coburn2025meta}. 

\item Speaking of RISC-V cores, Tenstorrent has released their second and third iteration of their RISC-V based accelerators, the Wormhole and Blackhole. The Wormhole accelerator is comprised of 80 Tensix processor, each of which is comprised of five RISC-V cores, for a total of 400 RISC-V cores. The Blackhole processor expands this to 140 Tensix processors, each also with five RISC-V cores, for a total of 700 RISC-V cores. Blackhole also includes 16 higher performance RISC-V cores for on-device hosting and running Linux and another 52 smaller RISC-V cores for memory management, communications, and system management~\cite{mann2024tenstorrent}. 

\item Based on take-aways of the DARPA Synapse neuromorphic commputing project from a decade ago, which included the development of the IBM TrueNorth accelerator~\cite{feldman2016ibm,esser2016convolutional,akopyan2015truenorth}, IBM has been developing and released the NorthPole inference accelerator~\cite{cassidy2024ibm,akopyan2024breakthrough,appuswamy2024breakthrough}. It features 256 cores that can execute 8-bit, 4-bit, and 2-bit digital arithmetic. It also has four on-chip networks to minimize communication hotspots. IBM has also announced that they will be releasing their AI Acceleration Unit (AIU) in 2026 as a PCIe card called Spyre~\cite{morgan2024ibm,berry2024ibm}. 

\item Japanese company Preferred Networks has released its second generation MN-Core2 chip, of which eight are integrated on each PCIe card~\cite{preferred2020mncore,makino2024mn-core2}. The accelerator is aimed at training in that it supports FP16, FP32, and FP64. Future versions will have an inference-focused version and a training-focused version. 

\item China's Huawei released two new versions of the Ascend 910: the 910B and 910C~\cite{morgan2024huawei,shilov2025huawei}. The original 910 has been labeled the 910A. These new Ascend accelerators have the same core design of the 910A, but they are now fabbed indigenously by SMIC (rather than TSMC in Taiwan due to American export restrictions). 

\item China's Cambricon, known mainly for its Kiren smartphone GPUs, has also recently released data center GPUs including the MLU290-M5 and MLU370-X8~\cite{shilov2025chinas}. 

\item Moore Threads, another Chinese GPU company, has emerged with a series of GPUs that can be used for business computers, gaming, and AI inference~\cite{klotz2024new}.

\item HyperX Logic (formerly Coherent Logix) previous generation HX40416 accelerator features 416 processing elements in a mesh topology, and each PE can execute 4 multiply-accumulate operations per clock cycle~\cite{demler2020coherent}. HyperX Logic has been focused on space applications along with  audio and video production with its programmable dataflow architecture, and has added AI inference applications to its application portfolio in recent years. 

\item The d-Matrix Corsair accelerator features arithmetic in SRAM memory cores and RISC-V control CPUs~\cite{ward2025d-matrix}. Each Corsair chiplet has 256 64-by-64 SRAM-arithmetic cell array cores, and there are four chiplets per Corsair package. A Corsair PCIe card has two Corsair packages, which totals 2048 cell array cores. This accelerator is aimed at small-batch, low-latency data center inference. 

\item The FuriosaAI RNGD implements tensor contractions as a computational primitive for AI inference~\cite{kim2024tcp}. Each Tensor Unit has 64 slices, and each slice has a tensor engine, vector engine and transpose engine. The accelerator is coupled to HBM memory for high bandwidth memory access for executing inference on very large models. 

\item Taiwan startup Neuchips introduced their Raptor N3000 AI accelerator chip, which is featured in their first product, the Evo accelerator PCIe card. The Raptor chip includes 10 matrix engines, two vector engines, and an embedding engine~\cite{neuchips2025raptor}. 

\item South Korean startup Rebellions released their ATOM Max accelerator for data center inference~\cite{ward2024rebellions,hong2025performance}. 

\item Finally, Syntiant has expanded its very-low power analog accelerator offerings with their NDP250 chip and at audio processing and wake word detection~\cite{ward2024syntiant}. 

\end{itemize}

Attrition among AI startups and corporate efforts continue to be part of normal business patterns. These accelerators have been removed from the survey because their accelerator(s) are no longer commercially available. Cornami has been removed because the company has moved their focus to embedded computing solutions for homomorphic encryption. 
AImotive has been removed because their accelerator product line is an RTL specification rather than a chip product. Three companies have been acquired or are closing down. The staff of Untether AI has been acquired by AMD, and the company is in the process of winding down~\cite{ward2025untether}, while Esperanto is also winding down with almost all of their employees having found other opportunities~\cite{ward2025esperanto}. Further, AlphaICs does not appear to be in business anymore, so their entry has been removed. 
Finally, after delivering the impressive Aurora supercomputer at Argonne National Laboratory, Intel cancelled its Xe-HPC (codenamed Ponte Vecchio) and plans for other high end computational GPUs.

We are also anxiously awaiting more details about several accelerators that have been announced, including peak performance and peak power numbers. Among the major American GPU vendors, Nvidia has released the names of the next two generations of GPUs, the Ruben and Feynman GPUs, while AMD has announced the expected release their MI430X and MI450X in the first half of 2026. Similarly, Baidu has announced its third generation P800 Kunlun accelerator~\cite{pan2025chinas}, while there is discussion that Huawei will be releasing Ascend 920 and 920C accelerators soon~\cite{morgan2025separate}. HyperX Logic has released a new space-focused accelerator called Midnight, for which we are hoping to see performance and power numbers. Horizon Robotics, which specializes in automotive inference accelerators has released Journey 5 and Journey 6 accelerators, but peak performance and power numbers are not yet available. 
Tesla has announced their Full Self-Driving Computer V2 (Gen5), but no details have been published yet. Finally, Q.ANT has released an optical accelerator, which is able to compute entire inference chains in the optical domain. This first version demonstrates the capability and opportunity of computing in the optical domain; their roadmap expects to meet and exceed the computational performance and energy efficiency of CMOS-based accelerators within the next few years.

\section{Observations and Trends}

\begin{figure*}[!bth]
    \centering
    \includegraphics[width=\textwidth]{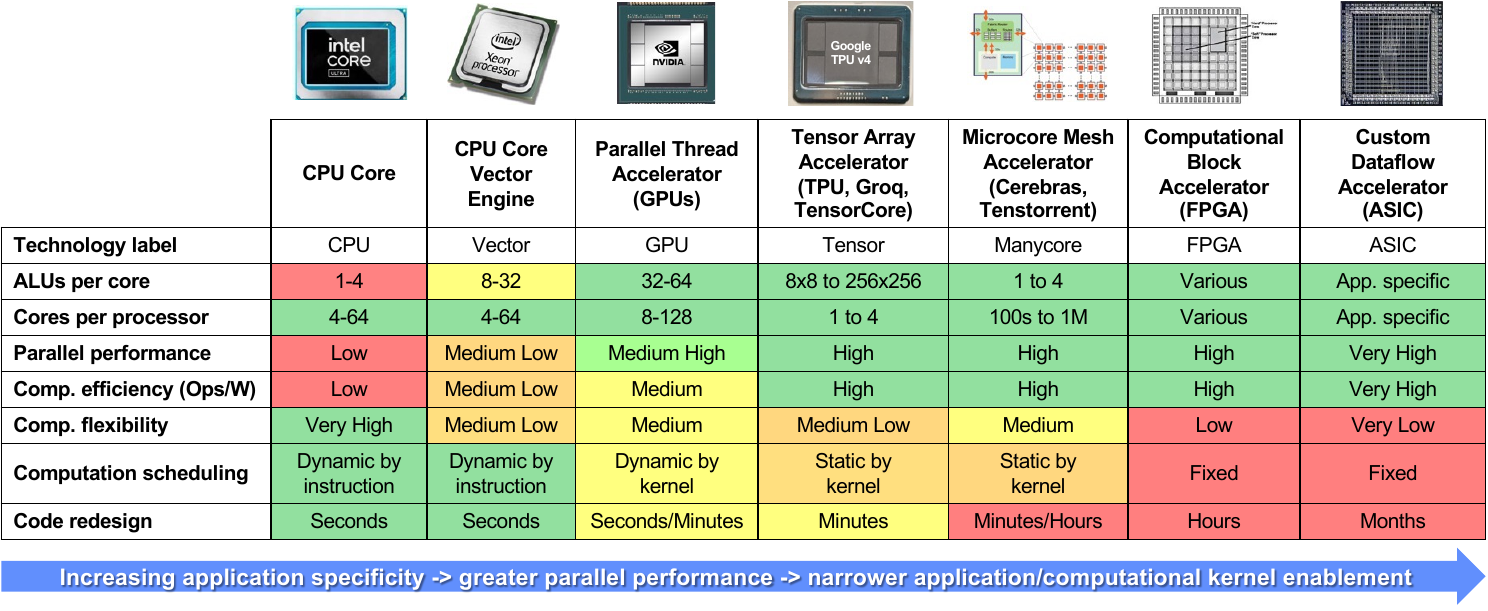}
    \caption{Range of AI accelerator computer architecture categories. Going from left to right, greater optimization of data movement between computations means data travels less distances between computations, and more computations executed in parallel. However, it also means less flexibility in operation types and programmability. CPU = Central Processing Unit; AVX = Advanced Vector eXtensions; SVE = Scalable Vector Extensions; GPU = Graphics Processing Unit; TPU = Tensor Processing Unit; CGRA = Course Grained Reconfigurable Architecture
FPGA = Field Programmable Gate Array; ASIC = Application Specific Integrated Circuit}
    \label{fig:architecture_comparison}
\end{figure*}

In the past two years since the last iteration of this survey, more details have been published in conference papers, journal papers, and technical press articles that have made it possible to more accurately generalize the main categories of architectures that are being used for AI processors/accelerators. These are summarized in Figure~\ref{fig:architecture_comparison}, and the categories span from highly flexible CPUs on the left to statically deployed FPGAs and ASICs on the left. 

While CPUs are very flexible and can execute all applications, they are not optimized to execute the highly parallel computations of inference and training. CPU vector engines are better suited for inference and training than CPUs, but they generally do not have memory coalescing features to pack strided memory accesses into dense vector accesses, which affects their parallel computational efficiency. 

At the other end of the categories are FPGAs and ASICs. We do not see ASICs much because of the requirement for most AI applications to support multiple different AI models with the same compute hardware. Similarly, FPGAs are used in embedded applications when AI models have been chosen for deployment, and some FPGAs even have tensor accelerator blocks included in the available computational blocks. 

The bulk of high performance and efficient accelerators are parallel thread accelerators (GPUs), tensor array accelerators, and microcore mesh accelerators because they are designed specifically for high main memory bandwidth, highly parallel computation and highly parallel and complex data movement that are required for highly efficient and highly performing inference and training. Among these, parallel thread accelerators are the most flexibly of these parallel compute engines. They have memory coalescing capability along with parallel compute cores (Symmetric Multiprocessors – SMs - in Nvidia parlance) that are dynamically scheduled with compute kernels. Code redesign and recompilation of kernels is fairly quick. Tensor array accelerators microcore mesh accelerators are more statically scheduled in that one preloads the model parameters and code into the accelerator before executing inference or training on them. To change the model parameters or code, the new one needs to be loaded into the accelerator anew, which involves some latency. Kernel code redesign and recompilation involves both compilation and mapping of the model code and parameters onto the compute elements, which involves an optimization of code and resources thereby taking more time than just compiling the numerical kernel code. But this often results in more efficient execution and higher performance than parallel thread accelerators.  

There are several more observations and comments for us to appreciate on Figure~\ref{fig:PeakPerformancePower}. 
\begin{itemize}
\item Int8 continues to be the default numerical precision for embedded, autonomous and data center inference applications, and fp16/bf16 has become the default numerical precision for training. However, some favorable outcomes have come from training in fp8 and even fp4, particularly for generative AI models, which save both computational and data movement energy.  
\item In our re-evaluation and re-categorization of each of the accelerators in this survey, we were pleasantly surprised by the variety of architectural choices being made to experiment, find, and exploit competitive advantages for using their accelerator in certain applications and for certain models. This was predicted in Nowatzki, et al.~\cite{nowatzki2017domain}, and it is encouraging to see it play out in commercial competition. 
\item In the data center domain, Nvidia continues to dominate the media coverage and sales for AI acceleration. However, AMD, Groq and Cerebras have become significant competitors to Nvidia, while many other commercial offerings are also gaining footholds. And it should be noted that the Groq accelerator is currently being manufactured using a 12-nm process and often outperforming accelerators at smaller circuit feature sizes, which supports the findings in~\cite{davies2025defying}. (Groq has announced a second generation accelerator that will be fabbed at a smaller circuit feature size.) And both Groq and Cerebras accelerators are not GPU architectures. 
\end{itemize}

\subsection{Non-CMOS Technologies}

As part of this survey, we continue to track other technologies that could be used to implement AI accelerators. Among them are memristors, neuromorphic architectures, cryogenic computing, and optical computing. In all of these domains, research and development continues to show opportunity and hope to become competitive with current commercial offerings. The one new development that is worth noting is in the optical computing area. While we still wait for the release of a commercial accelerator from LightMatter, Lightelligence, and LightOn (we are assured that they are coming!), Q.ant has released a commercial optical accelerator, as we mentioned Section~\ref{sec:survey}. This is an exciting development, and we will be watching how they and other optical computing vendors compete among computational accelerators. 




\section{Summary}

This paper updates the Lincoln AI Computing Survey (LAICS) of AI accelerators that span from extremely low power through embedded and autonomous applications to data center class accelerators for inference and training. We presented the new full scatter plot along with zoomed in scatter plots for each of the major deployment/market segments, and we discussed the new additions for the year. We also included a categorization of AI computing hardware based on much new material that has been published for these AI accelerators. 

\section{Data Availability}

The data spreadsheets and references that have been collected for this study and its papers are posted at \url{https://github.com/areuther/ai-accelerators} after they have cleared the release review process. 

\section*{Acknowledgement}

We express our gratitude to 
LaToya Anderson, Masahiro Arakawa, Bill Arcand, Bill Bergeron, David Bestor, Bob Bond, Alex Bonn, Chansup Byun, Vitaliy Gleyzer, Jeff Gottschalk, Michael Houle, Matthew Hubbell, Hayden Jananthan, David Martinez, Lauren Milechin, Sanjeev Mohindra, Paul Monticciolo, Julie Mullen, Andrew Prout, Stephan Rejto, Antonio Rosa, Charles Yee, and Marc Zissman
for their support of this work. We are also grateful to Mark Gouker, Bob Atkins, and Livia Racz for the discussions that eventually were captured in the accelerator categorizations. 


\bibliographystyle{IEEEtran} 
\bibliography{LAICSAcceleratorTrends2025}


\end{document}